\documentclass[amsmath,amssymb,aps,prd,twocolumn,english]{revtex4-2}
\usepackage{graphicx}
\usepackage{amsmath}
\usepackage{amssymb}
\usepackage[colorlinks=true,citecolor=blue,linkcolor=blue,urlcolor=blue]{hyperref}
\usepackage{cleveref}
\usepackage{tabularx}
\usepackage{placeins}
\newcommand{\be}{\begin{equation}}
\newcommand{\ee}{\end{equation}}
\begin{document}

\title{Atomic probe of dark matter differential interactions with subatomic particles}

\author{Yossi Rosenzweig}
\thanks{These two authors contributed equally.}
\affiliation{Department of Physics, Ben-Gurion University of the Negev, Israel}

\author{Yevgeny Kats}
\thanks{These two authors contributed equally.}
\affiliation{Department of Physics, Ben-Gurion University of the Negev, Israel}

\author{Menachem Givon}
\affiliation{Department of Physics, Ben-Gurion University of the Negev, Israel}

\author{Yonathan Japha}
\affiliation{Department of Physics, Ben-Gurion University of the Negev, Israel}

\author{Ron Folman}
\affiliation{Department of Physics, Ben-Gurion University of the Negev, Israel}

\begin{abstract}
Searching for physics beyond the Standard Model is one of the main tasks of experimental physics. Candidates for dark matter include axionlike ultralight bosonic particles. Comagnetometers form ultrahigh sensitivity probes for such particles and any exotic field that interacts with the spin of an atom. Here, we propose a multiatom-species probe that enables one not only to discover such fields and measure their spectrum but also to determine the ratios of their coupling strengths to the subatomic particles---electrons, neutrons, and protons. We further show that the multifaceted capabilities of this probe can be demonstrated with synthetic exotic fields generated by a combination of regular magnetic fields and light-induced fictitious magnetic fields in alkali atoms.  These synthetic fields also enable the accurate calibration of any magnetometer or comagnetometer probe for exotic physics.
\end{abstract}
\maketitle

\section{Introduction}
Searching for physics beyond the Standard Model (BSM) is one of the main tasks of experimental physics if we are to explain yet ill-understood phenomena such as the strong $CP$ problem\,\cite{peccei1977cp}, neutrino oscillations\,\cite{ahmad2001measurement,barger2012physics}, matter-antimatter asymmetry\,\cite{canetti2012matter}, dark energy\,\cite{peebles2003cosmological,ade2014planck}, dark matter (DM)\,\cite{bertone2018history,kimball2022search}, and the period of cosmic inflation\,\cite{Guth:1980zm,Starobinsky:1980te,Linde:1981mu,Planck:2018jri} and what preceded it. Although it is natural to associate fundamental and BSM experiments with a large-scale facility such as the Large Hadron Collider at CERN or the Super-Kamiokande Cherenkov detector, many such experiments can be done using a tabletop experiment\,\cite{demille2017probing}. One such experimental tool is the spin-sensitive comagnetometer\,\cite{kornack2002dynamics}. As will be explained in detail later on, the comagnetometer attenuates low-frequency magnetic fields while keeping its sensitivity to exotic fields spin interaction, making it an excellent tool for BSM searches\,\cite{smiciklas2011new,brown2010new,terrano2021comagnetometer}. One class of BSM scenarios that can be probed with comagnetometers is dark matter in the form of axionlike particles (ALPs)\,\cite{ParticleDataGroup:2022pth}. These are light spin-0 particles that arise generically as pseudo-Nambu-Goldstone bosons of spontaneously broken global symmetries. They are commonly long-lived and could be produced during the cosmological evolution of the Universe in the right amount to account for the observed DM abundance. For ALP masses $m_a\ll 1$\,eV, these ultralight bosonic particles will have high occupation numbers and behave as a classical field oscillating with frequency $f \simeq m_a/(2\pi)$ (where we use natural units with $\hbar=c=1$). One of the strong motivations for ALPs is the QCD axion---a hypothetical particle naturally emerging from a solution to the strong $CP$ problem\,\cite{peccei1977cp,weinberg1978new}. In the simplest scenarios for ALP DM production, via the misalignment mechanism, the QCD axion mass needs to be $m_a \gtrsim \mu$eV for consistency with the observed amount of DM\,\cite{Adams:2022pbo}, and such masses cannot be probed with comagnetometers. However, there exist scenarios in which a much lighter QCD axion can naturally produce the observed DM abundance\,\cite{DiLuzio:2021gos,Papageorgiou:2022prc,Choi:2022nlt}.

There are several ways for ALPs to interact with the Standard Model particles\,\cite{safronova2018search, kimball2022search}. In particular, they would, in many cases, couple to Standard Model fermion fields $\psi$ via
\begin{equation}
    {\cal L} = \frac{g_f}{f_a}\, \partial_\mu a\,\bar \psi\gamma^\mu\gamma^5\psi \,,
\end{equation}
where $a$ is the ALP field, $f_a$ is the spontaneous symmetry-breaking energy scale, and $g_f$ is a model-dependent and fermion-dependent coupling factor (analogous to charge and dimensionless in the natural units). In the nonrelativistic limit, one obtains an interaction analogous to that of a spin with a magnetic field\,\cite{Graham:2013gfa}, with the Hamiltonian:
\begin{eqnarray}
\mathcal{H}\approx-\frac{g_f}{f_a}\frac{\mathbf{S}}{\lvert S \rvert}\cdot\nabla a\,,
\label{eq: exotic Hamiltonian}
\end{eqnarray}
where $\mathbf{S}$ is the fermion spin ($|S|=1/2$ is the maximum spin projection), and $\nabla a$ is the spatial gradient of the ALP field. While in the simplest QCD axion models the couplings $g_f$ are commonly of order one, which would make the effect extremely small, there exist scenarios in which they are enhanced by orders of magnitude\,\cite{Darme:2020gyx}. There are several ongoing experiments with the goal of detecting such an interaction\,\cite{jiang2021search,JacksonKimball:2017elr,pospelov2013detecting,masia2020analysis,Abel:2017rtm,Bloch:2022kjm}.

The ALP couplings to electrons and nucleons are subject to constraints from star cooling\,\cite{Capozzi:2020cbu} and neutron star cooling\,\cite{Buschmann:2021juv}, respectively. These bound the possible signal size from a galactic ALP DM halo, making it challenging (although not impossible\,\cite{Lee:2022vvb,Wei:2023rzs}) for comagnetometers to observe. However, if the ALP DM is not smooth but forms ALP stars that the Earth encounters once in a while\,\cite{Braaten:2019knj,JacksonKimball:2017qgk}, or produces a halo around the Sun, its signals can be detectable also for smaller couplings. Domain walls of the ALP field are another possible target for comagnetometers. These and additional possibilities are reviewed in Ref.\,\cite{GNOME:2023rpz}. While our following derivation focuses on a steady-state (i.e., single frequency) exotic perturbation, the method also applies to the aforementioned transient events, which exhibit a rich frequency domain.

\begin{figure*}
\includegraphics[width=1.01\columnwidth]{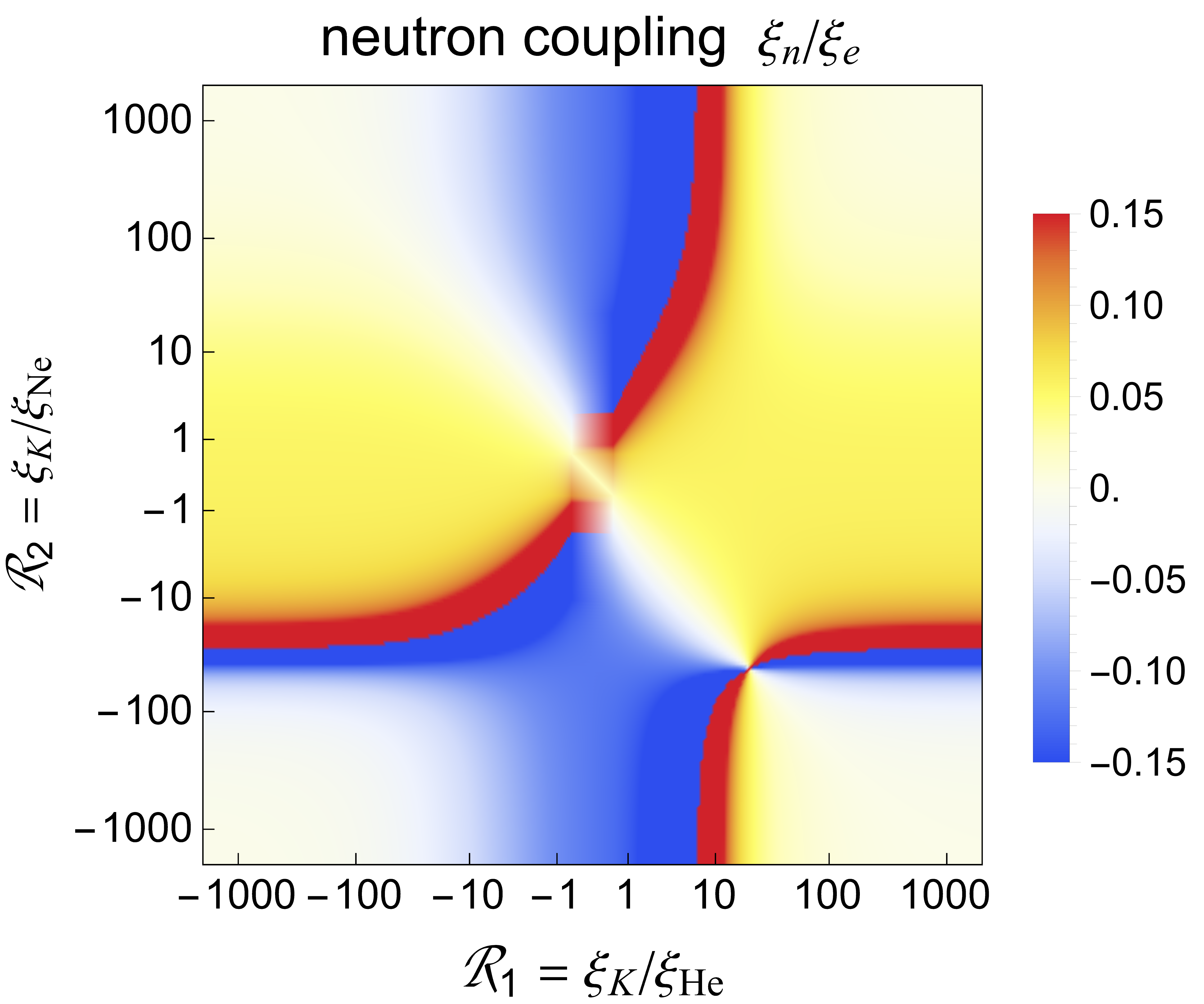}\qquad
\includegraphics[width=0.96\columnwidth]{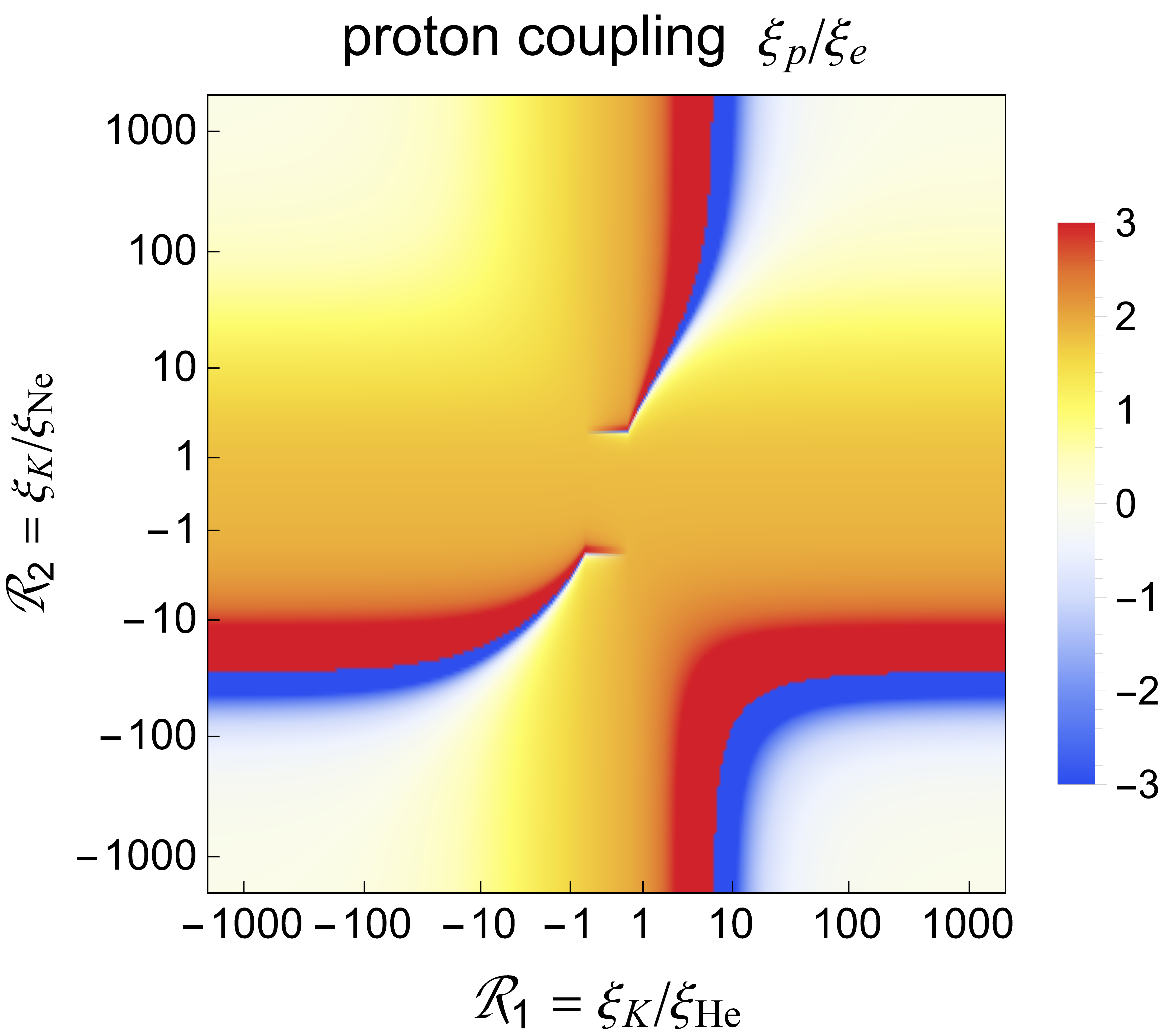}
\caption{Coupling constants of the ALP field with neutrons $\xi_n$ (left) and protons $\xi_p$ (right) in units of the coupling with electrons $\xi_e$ deduced from two atomic coupling ratios $\mathcal{R}_N=\xi_A/\xi_N$ with the same alkali atom ($A=\,^{39}$K) and two different noble-gas atoms ($N=\,^{3}$He, $^{21}$Ne). Each atomic coupling ratio is obtained from simultaneous measurements in two comagnetometers with the same atomic species but different operation parameters. 
The values of the subatomic particle coupling constants are calculated using nuclear spin contents for the alkali and noble-gas atoms, as given in Table~\ref{tab:nuclear-spin}. Most of the points in the $\mathcal{R}_1$--$\mathcal{R}_2$ plane require the coupling to protons to be larger than to neutrons because the proton spin fractions in both noble gases according to the models used for the spin contents are small.} 
\label{fig:xi2D}
\end{figure*}

\section{The apparatus}
Comagnetometers are gaining popularity in testing ALP-fermion interaction theories\,\cite{terrano2021comagnetometer,bloch2020axion,wu2019search,JacksonKimball:2023snl,Lee:2022vvb,Wei:2023rzs}. We focus on alkali-noble-gas comagnetometers due to their relative simplicity and high sensitivity to spin interactions with exotic fields, while having an attenuated response to low-frequency magnetic fields\,\cite{padniuk2022response,PhysRevApplied.19.044092}.
A comagnetometer of this type comprises an alkali vapor and a noble gas with a nonzero nuclear spin confined in a glass cell. The cell is heated to have a high vapor density and placed inside a magnetic shield. A pump laser polarizes the alkali vapor along the axis of the beam (z axis), which polarizes the noble gas along the same axis via spin-exchange collisions\,\cite{walker1997spin}.
A linearly polarized probe laser beam along the x axis measures the spin projection along this axis. If there were no noble gas, a magnetic field along the y axis would cause the spin to precess in the x-z plane and be detected by the probe beam\,\cite{savukov2017spin}. When the polarized noble gas is present, the alkali atoms are affected by the effective magnetic field\,\cite{Schaefer-et-al} it induces and vice versa. The induced fields are countered by applying an external magnetic field equal in amplitude but with an opposite sign (``compensation field")\,\cite{kornack2002dynamics}. As a result, the polarized noble gas will follow a slowly varying transverse magnetic field (with respect to the pump and probe axes), causing the response of the alkali to remain (close to) zero, making the comagnetometer immune to low-frequency magnetic field noise. This makes the comagnetometer a leading probe for BSM spin interactions, as it attenuates the low-frequency response to regular magnetic fields but not to pseudomagnetic exotic fields that presumably interact with spins via different couplings.
\par
Unlike the magnetic field, which interacts with the electron and nucleon spins with coupling strengths given by their magnetic moments, the ALP exotic field can couple to the electron, proton, and neutron spins with unknown arbitrary strengths. Coupling to the electronic spin affects only the alkali atoms, which have nonzero electronic spin, while coupling to nuclear spin would mainly affect the magnetization in the vapor cell due to noble gas in the comagnetometer, whose density is much larger than that of the alkali gas. Furthermore, coupling to the nucleons will also affect the spin of the alkali through the large magnetic field induced by the polarized noble gas on the alkali. An exotic field whose ratio between the coupling to the electron and nucleon spins is significantly different from that of the magnetic field, would evade the compensation effect and lead to a strong comagnetometer response. Importantly, to avoid attenuation of exotic fields coupled to the electron spin, the magnetic shielding should be based on electric current (conductors or superconductors) rather than spin (soft ferromagnets or ferrimagnets)\,\cite{kimball2016magnetic}.
\par
While detecting a signal related to ALPs can reveal the field's existence, it cannot determine its amplitude because the coupling of the field to the atoms is not known. It is also impossible to determine the coupling constants of the exotic field with the subatomic particles or the coupling to each of the atomic species. Here, we propose a scheme that uses simultaneous detection in a few comagnetometers, some of them with different atomic species, to determine the ratios of coupling constants to the subatomic particles. While the proposed scheme will not increase the individual comagnetometers sensitivity to exotic fields, it can play a crucial role in characterizing the nature of the exotic field once it is detected. We further show that this scheme can be readily demonstrated with the help of a synthetic exotic field.
\par
\begin{figure}
\centering
\includegraphics[width=\linewidth]{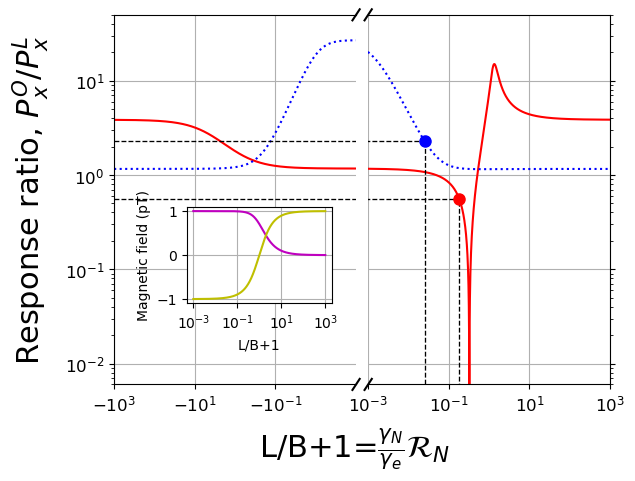}
\caption{Response ratios of different comagnetometers (with parameter values listed in Table~\ref{tab:parameter_table}) to a synthetic exotic field perturbation at $1\,$Hz as a function of the ratio between the light shift ($L$) and magnetic ($B$) fields in the system. The latter ratio is related to the ratio of exotic field coupling to the alkali and noble-gas spin as $L/B+1 = \frac{\gamma_N}{\gamma_e}\mathcal{R}_N$ [see Eq.\,\eqref{eq:bB,bL}]. The response ratios of $^{39}$K-$^{3}$He and $^{39}$K-$^{21}$Ne comagnetometers with optimal pumping rate $P^O_x$ to similar comagnetometers with a low pumping rate $P^L_x$ are shown in solid red and dotted blue, respectively. The inset shows how each value of $\mathcal{R}_N$ can be synthetically generated by a combination of a light shift and magnetic field (yellow and magenta lines, respectively). The dashed black line shows a possible $\mathcal{R}_N$ measurement in each pair of comagnetometers.
While the same value of a response ratio can usually be obtained at two different points on the curve, this apparent ambiguity can be resolved by looking at the response ratio of a pair of comagnetometers with different noble gases (see Table~\ref{tab:ambiguity}).}
\label{fig: comag response with B and L}
\end{figure}

In addition, calibration is a major problem in BSM searches since we cannot generate a calibration signal of an exotic field (otherwise, it would not be ``exotic''). While a typical magnetometer can be tested by sending a pulse with known amplitude and comparing the results to the known value, this type of procedure is less relevant in BSM sensors. Recently, it was suggested to calibrate a comagnetometer by comparison of a detailed theoretical model to frequency-dependent measurements of a response to a transient magnetic field\,\cite{padniuk2023universal}. Here, however, we do not send a magnetic pulse but suggest generating a synthetic exotic field with a different coupling ratio than a magnetic field. We describe how to generate a synthetic exotic field by a combination of a magnetic field and an ac Stark shift field (later on, in Fig.\,\ref{fig: comag response with B and L}, we show actual experimental parameters that generate such fields) and focus on the steady-state response. The combination of the fields mimics the effect of exotic fields and, therefore, can be used to calibrate comagnetometers by directly measuring their response to such fields.

\begin{table}
\centering
\renewcommand{\arraystretch}{1.3} 
\begin{tabular}{cccc}
\hline\hline
Nucleus & $\sigma_n$ & $\sigma_p$ & Refs. \\
\hline
$^3$He    & $0.87$  &      $-0.027$ & \cite{kimball2015nuclear,Friar:1990vx} \\
$^{21}$Ne & $0.196$ & \;\;\,$0.013$ & \cite{Engel:1989ix} \\
$^{39}$K  & $0.034$ &      $-0.131$ & \cite{kimball2015nuclear,Engel:1995gw} \\
\hline\hline
\end{tabular}
\caption{Fractions of the nuclear spin due to neutron and proton spins. While the $^3$He numbers are based on a full-scale shell model calculation~\cite{Friar:1990vx} and the $^{39}$K numbers are based on a detailed perturbation theory calculation~\cite{Engel:1995gw}, the numbers for $^{21}$Ne are estimates based on mirror nucleus properties~\cite{Engel:1989ix} and might not be as accurate. Very similar numbers for $^{21}$Ne were obtained in~\cite{Brown:2016ipd} using the configuration interaction shell model, but other models lead to different predictions\,\cite{Stadnik:2014xja}.}
\label{tab:nuclear-spin}
\end{table}

\par
We begin by presenting in Fig.\,\ref{fig:xi2D} an example of the output of the new probe, namely, the determination of the ratios of the different fundamental couplings between the exotic field and the subatomic particles. The coupling constants $\xi_f\equiv g_f/f_a$ of the exotic field to the subatomic fermions ($f=e,n,p$ for electrons, neutrons and protons) are not known. Alas, a single measurement cannot reveal these constants because each comagnetometer uses two atomic species: alkali atoms and noble-gas atoms and the coupling of the exotic field to each of the atomic species is a weighted sum over its couplings to the subatomic particles. In particular, the coupling to the nucleus of a given atom is given by $\xi_{\rm nuc}=\sigma_n\xi_n+\sigma_p\xi_p$, where $\sigma_n$ and $\sigma_p$ are the fractional contributions due to neutron and proton spins to the spin of the specific nucleus\,\cite{kimball2015nuclear}. In our scheme, we first extract the ratio $\mathcal{R}_N\equiv \xi_A/\xi_N$ of the coupling of the exotic field with the alkali atoms ($A$) and the noble-gas atoms ($N$) used in a pair of comagnetometers with the same atomic constituents but different operation parameters. Then, by taking the results of such a ratio from two pairs of comagnetometers using different atomic species we can also extract the coupling ratios $\xi_n/\xi_e$ and $\xi_p/\xi_e$ of the subatomic particles, as demonstrated in Fig.\,\ref{fig:xi2D}.  

\section{Theoretical description}
Let us now detail how a synthetic field enables one to demonstrate, calibrate, and optimize the comagnetometer response to ALP-electronic/nuclear spin interaction, and utilize this understanding to explain how the suggested novel probe enables the measurement of the fundamental couplings. Light near the alkali optical resonance results in an ac Stark shift, which manifests itself as a fictitious magnetic field via the vector polarizability\,\cite{le2013dynamical}, while this field does not exist for noble gas, as the latter has only scalar polarizability\,\cite{dzuba2018screening}.
This makes the light-induced fictitious magnetic field indistinguishable from an exotic field that couples only to the alkali spin. Generating a synthetic field that couples only to the nuclei of the noble gas is also possible in a comagnetometer by applying a magnetic field and compensating its effect on the alkali by generating an equal and opposite fictitious magnetic field. This way, the magnetic field affects only the polarized noble gas, mimicking the effect of exotic field coupling only to the
noble-gas nucleus. A general model with an arbitrary ratio between the coupling factors of the exotic field to the spins of the alkali and the noble gas can be tested using a combination of regular and fictitious magnetic fields.
\par
The dynamics due to the interaction of the spins of the gases ($n_g$ species) with the magnetic, exotic and optical fields and the interaction between the spins is described by the set of Bloch equations for the polarizations $\mathbf{P}_j$ ($1\leq j \leq n_g$) of the electron spins in the alkali atoms or the nuclear spins in the noble-gas atoms
\begin{gather}
\dot{\mathbf{P}}_j = \frac{1}{q_j}\left\{\left[\gamma_j\left(\mathbf{B}+\sum_k\lambda_{jk}M_0^k\mathbf{P}_k+\mathbf{L}_j\right)+\xi_j\mathbf{b}\right]\times \mathbf{P}_j \right.  \nonumber \\
\quad\;\;\left. + \sum_k \kappa_{jk}n_k(\mathbf{P}_k-\mathbf{P}_j)-R_{pj}(\mathbf{P}_j-\mathbf{s})-\Gamma_j\mathbf{P}_j\right\} , \label{Eq: Bloch}
\end{gather}
where we sum over all existing species (index $k$). $\mathbf{P}_j$ are normalized to a maximal length of unity, $\bf{B}$ is the magnetic field vector, and $\mathbf{L}_j$ is the fictitious magnetic field (ac Stark shift) due to the interaction between the laser light field and the atoms.
Here $\gamma_j$ are the gyromagnetic ratios: electron gyromagnetic ratio $\gamma_e$ for the alkali atoms, and $\gamma_N$ for the nuclear spin of the noble gas, and $q_j$ are the nuclear slowing-down factors for the alkali atoms, and $1$ for the noble gas.
The pseudomagnetic field due to an exotic field interaction $\mathbf{b}$ is proportional to the ALP field gradient $\nabla a$. It couples to the atoms via the constants $\xi_j$ representing the coupling to the electronic and nuclear spins of each species. These coupling factors emerge from the coupling strengths $g_e/f_a$, $g_p/f_a$, and $g_n/f_a$ for the exotic field interaction with the spin of the electrons, protons, and neutrons in the atoms\,\cite{kimball2015nuclear}.
In addition to the coupling to external fields, each species $j$ experiences an effective magnetic field\,\cite{Schaefer-et-al} induced by the magnetization of each of the other species $k$ through the coupling factors $\lambda_{jk}$, which include a geometrical and temperature-dependent enhancement factor. The magnetization is proportional to the normalized polarization $\mathbf{P}_k$ and the maximal magnetization $M_0^k=\mu_k n_k$ where $n_k$ is the density and $\mu_k$ is the magnetic moment of the atom $k$.
The second line of Eq.\,\eqref{Eq: Bloch} includes the incoherent processes: laser pumping with a rate $R_{pj}$
and direction unit vector $\mathbf{s}$, spin exchange with a rate proportional to $\kappa_{jk}$\,\cite{babcock2003hybrid}, and $\Gamma_{j}$ accounts for all the remaining relaxation processes for species $j$. The parameter values we use are listed in Table~\ref{tab:parameter_table}, and a method for obtaining the steady-state solution of the coupled Bloch equations in response to a monochromatic exotic field is described in the Appendix.

\begin{table*}
    \centering
    \renewcommand{\arraystretch}{1.3} 
    \begin{tabular}{lcc}
        \hline\hline
        Parameter & Value & Units \\
        \hline
        Temperature, $T$ & 190 & $^\circ$C \\
        K-$^3$He spin-exchange cross section rate, $\kappa_{\mathrm{K-He}}$ & $5.5\cdot10^{-20}$ & cm$^3$/s \\
        K-$^{21}$Ne spin-exchange cross section rate, $\kappa_{\mathrm{K-Ne}}$ & $3\cdot10^{-20}$ & cm$^3$/s \\
        K-$^3$He enhancement factor, $\lambda_{\mathrm{K-He}}$ & $8\pi/3\times5.9$ &  \\
        K-$^{21}$Ne enhancement factor, $\lambda_{\mathrm{K-Ne}}$ & $8\pi/3\times30$ &  \\
        K concentration, $n_\mathrm{K}$ & $9.7\cdot10^{13}$ & cm$^{-3}$ \\
        $^3$He concentration, $n_{\mathrm{He}}$ & $3.5$ & amg \\
        $^{21}$Ne concentration, $n_{\mathrm{Ne}}$ & $3.5$ & amg \\
        Electron gyromagnetic ratio, $\gamma_e$ & $2\pi\cdot2.8$ & MHz/G \\
        $^{3}$He nuclear gyromagnetic ratio, $\gamma_\mathrm{He}$ & $2\pi\cdot3.24$ & kHz/G \\
        $^{21}$Ne nuclear gyromagnetic ratio, $\gamma_\mathrm{Ne}$ & $2\pi\cdot0.337$ & kHz/G \\
        K magnetic moment, $\mu_\mathrm{K}$ & $9.274\cdot10^{-21}$ & erg/G \\
        $^{3}$He magnetic moment, $\mu_\mathrm{He}$ & $1.154\cdot10^{-3}\mu_\mathrm{K}$ & erg/G \\
        $^{21}$Ne magnetic moment, $\mu_\mathrm{Ne}$ & $0.359\cdot10^{-3}\mu_\mathrm{K}$ & erg/G \\
        K relaxation rate, $\Gamma_\mathrm{K}$ & 600 & Hz \\
        $^3$He relaxation rate, $\Gamma_{\mathrm{He}}$ & $5\cdot10^{-5}$ & Hz \\
        $^{21}$Ne relaxation rate, $\Gamma_{\mathrm{Ne}}$ & $30\cdot10^{-5}$ & Hz \\
        Optimal pumping rate, $R_{p,\mathrm{K}}$ & $600$ & Hz \\
        Low pumping rate, $R_{p,\mathrm{K}}$ & $50$ & Hz \\
        K nuclear slowing-down factor, $q_\mathrm{K}$ & 5.2 &  \\
        \hline\hline
    \end{tabular}
    \caption{Simulation parameters.}
    \label{tab:parameter_table}
\end{table*}

\begin{figure}
\centering
\includegraphics[width=0.95\linewidth]{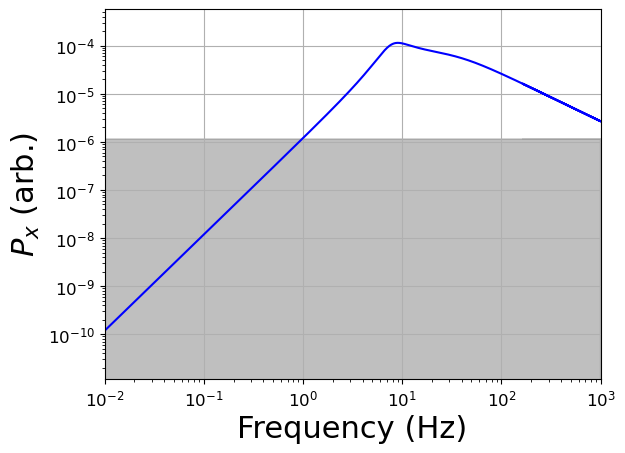}
\caption{Response of a $^{39}$K-$^3$He comagnetometer (for parameter values listed in Table~\ref{tab:parameter_table}) to a (weak) 1\,pT magnetic field oscillating with frequency $f$. The magnetic field perturbation is perpendicular to the pump and probe beams. The response is the steady-state solution of Eq.\,\eqref{Eq: Bloch} for the x component of the polarization of the probed alkali. The gray area represents polarization values below the minimal detectable value, which was taken to be the maximal polarization generated by an oscillating magnetic signal of 10\,fT at 1\,Hz. It can be seen that low-frequency magnetic fields are attenuated as expected from a comagnetometer.}
\label{fig: comag frequency response}
\end{figure}

Figure~\ref{fig: comag frequency response} presents the comagnetometer steady-state response to a transverse (with respect to the pump and probe axes) magnetic field perturbation as a function of the perturbation frequency. We see in Fig.\,\ref{fig: comag frequency response} that at low frequencies the response is attenuated by the mechanism described above. The same response is expected for a pseudomagnetic exotic field if $\xi_A/\xi_N=\gamma_e/\gamma_N$, such that the exotic field mimics a magnetic field. However, in the general case the two coupling constants may have an arbitrary ratio $\mathcal{R}_N \equiv \xi_A/\xi_N$ and the response is then expected to usually be much stronger if $\mathcal{R}_N\neq \gamma_e/\gamma_N$.

\begin{figure}
\centering
\includegraphics[width=0.96\linewidth]{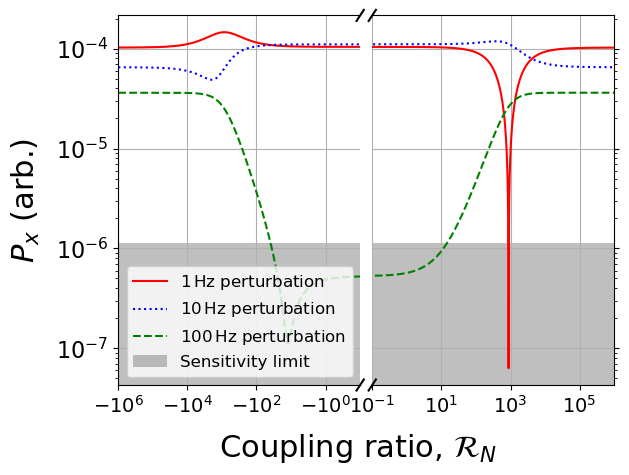}
\caption{Response of a $^{39}$K-$^3$He comagnetometer (for parameter values listed in Table~\ref{tab:parameter_table}) to exotic field perturbations of different frequencies as a function of the ratio between its coupling to the alkali and noble gas $\mathcal{R}_N=\xi_A/\xi_N$.
The magnitude of the effective pseudomagnetic fields $b_A\equiv \xi_A b/\gamma_e$ and $b_N\equiv \xi_Nb/\gamma_N$ corresponding to the coupling with the alkali and noble gas was normalized to be $\sqrt{b_A^2+b_N^2}=\sqrt{2}$\,pT for any value of $\mathcal{R}_N$. For $\mathcal{R}_N=\gamma_e/\gamma_N$, the exotic field is equivalent to a 1\,pT magnetic field.
The gray area represents the range of $P_x$ values below the minimal detectable value, which was taken to be the maximal polarization generated by an oscillating magnetic signal of 10\,fT at 1\,Hz.}
\label{fig: comag ratio response}
\end{figure}

In Fig.\,\ref{fig: comag ratio response}, we show the steady-state response of a comagnetometer to an exotic field perturbation (perpendicular to the pump and probe beams) as a function of the ratio $\mathcal{R}_N \equiv \xi_A/\xi_N$ between the exotic field coupling to the alkali and noble gas, for several oscillation frequencies. For an exotic field with $\mathcal{R}_N=\gamma_e/\gamma_N$, such that it mimics a magnetic field, the response at low frequencies is attenuated by the mechanism described above, while the response is generally much stronger if $\mathcal{R}_N\neq \gamma_e/\gamma_N$. It is interesting to note that the attenuation slightly above $\mathcal{R}_N = \gamma_e/\gamma_N$ is bigger than for exactly $\mathcal{R}_N=\gamma_e/\gamma_N$ as the signal due to the coupling to the electron is out of phase from the signal due to a magnetic-field-like combination of couplings\,\cite{padniuk2022response}, resulting in a destructive interference.

\section{The multiatom-species probe}
With the above understanding, we now show how the multiatom-species probe works. The interaction of ALP fields involves three unknown coupling parameters: $\xi_e$, $\xi_p$, and $\xi_n$ for the interaction with electrons, protons, and neutrons. A detection by a comagnetometer with given alkali and noble atoms may reveal the frequency of the exotic field but would teach us nothing about the amplitude of the field or the coupling strengths, which constitute four unknowns. For small perturbations, the response of the comagnetometer is proportional to the amplitude $b_y$ of the exotic field. Let us now consider two comagnetometers with different response curves $P^j_x(\mathcal{R}_j,b_y)$, where $P^j_x$ and $\mathcal{R}_j$ for $j=1,2$ represent the alkali polarizations and coupling ratios of the two comagnetometers. The ratio $P^1_x/P^2_x$ is independent of $b_y$. If the two comagnetometers have the same noble gas $N$ but different response curves due to different system parameters (such as gas pressures, pumping rates, etc.), the ratio of the responses reveals the value of the coupling ratio $\mathcal{R}_N$. The responses of the comagnetometers as a function of $\mathcal{R}_N$ can be experimentally simulated by a synthetic exotic field combined from the two applied fields $L_y$ and $B_y$: For any given value of $\xi_A$ (coupling to the alkali) and $\xi_N$ (coupling to the noble gas), the following perturbations along the y axis simulate an exotic field:
\be
    B_y = \frac{\xi_N}{\gamma_N}b_y \,,\qquad
    L_y = \left(\frac{\xi_A}{\gamma_e}-\frac{\xi_N}{\gamma_N}\right)b_y \,.
\label{eq:bB,bL}
\ee
The couplings $\xi_A$ and $\xi_N$ are linear combinations of the proton and neutron couplings, with coefficients that depend on the specific nucleus\,\cite{kimball2015nuclear}, and in the case of $\xi_A$ also the electron coupling. The ratio of the responses of two comagnetometers as a function of $\mathcal{R}_N$ is depicted in Fig.\,\ref{fig: comag response with B and L} for two pairs of comagnetometers with a different noble gas, such that the two comagnetometers in each pair have different response curves due to two different pumping rates: optimal and low (with a ratio of $10$).
Simultaneous measurement of an exotic field signal in the four comagnetometers enables the determination of the coupling ratios for the subatomic particles.
Measurement of $\mathcal{R}_{N_1}$ and $\mathcal{R}_{N_2}$ from the response ratios of two comagnetometers that share the same noble gas $N_1$ or $N_2$, respectively, may reveal the ratios of the exotic field couplings to the electrons ($\xi_e$), protons ($\xi_p$), and neutrons ($\xi_n$) using the following equations:
\begin{eqnarray}
\frac{\xi_e+(q_A-1)(\sigma^A_p\xi_p+\sigma^A_n\xi_n)}{\sigma^{N_i}_p\xi_p+\sigma^{N_i}_n\xi_n}&=& \mathcal{R}_{N_i} \,,\;\;
i = 1,2\,,
\label{eq: ratio}
\end{eqnarray}
where $\sigma_p$ and $\sigma_n$ are the fractions of the nuclear spin due to proton and neutron spins, respectively, in the alkali ($A$) or one of the noble gases ($N_{1,2}$), whose values are provided in Table~\ref{tab:nuclear-spin}.
Solving the set of two equations would reveal the ratios $g_p/g_e$ and $g_n/g_e$. In fact, a set of only three comagnetometers with one of them having a different noble gas is sufficient for determining these ratios, but the example of two pairs of comagnetometers is simpler to present.
In Fig.\,\ref{fig: comag response with B and L}, we show the ratios between the responses of two pairs of  comagnetometers as a function of the ratio between the fictitious magnetic field perturbation $L_y$ and the regular magnetic field perturbation $B_y$, which corresponds to the coupling ratios $\mathcal{R}_N$ of the comagnetometers. These response ratios enable the differential output of the probe, as presented in Fig.\,\ref{fig:xi2D}.

As an outlook, we note that the analysis in this work utilized the magnitude of the response. However, the phase of a comagnetometer response to an exotic field perturbation also carries information\,\cite{padniuk2022response} that can be utilized whether the two comagnetometers experience the same perturbation at the same time\,\cite{Masia-Roig:2022net} or with a time delay. An additional future improvement would come from work narrowing the uncertainties concerning proton and neutron spin fractions in the different elements\,\cite{kimball2015nuclear}. If theoretical models are not accurate enough, accuracy improvement can be achieved by utilizing additional noble-gas species with different spin contents (e.g., $^{129}$Xe).

\begin{table}
\centering
\renewcommand{\arraystretch}{1.3} 
\begin{tabular}{c|c|c}
\hline\hline
 & \,Red circle\, & \,Alternative\, \\\hline
$P_x^O$ & 2.47 & 1.31 \\
$P_x^L$ & 4.40 & 2.35 \\
\hline\hline
\end{tabular}
\caption{Responses (in arbitrary units) of the $^{39}$K-$^3$He comagnetometers from Fig.~\ref{fig: comag response with B and L} with the optimal pumping rate ($P_x^O$) and low pumping rate ($P_x^L$), for the point indicated by the red circle and the other point that gives the same ratio of responses. Since the values are different, their ratios with the response of any of the $^{39}$K-$^{21}$Ne comagnetometers will be different, which will allow for resolving the ambiguity.}
\label{tab:ambiguity}
\end{table}

\begin{acknowledgments}
We thankfully acknowledge useful discussions with Dmitry Budker, Derek F.\ Jackson Kimball, and Joshua Eby. This work was supported by the United States---Israel Binational Science Foundation Grants No.~2016635 and No.~2018257, the Israel Science Foundation Grant No.~1666/22, and in part by the Israel Innovation Authority Grants No.~67082 and No.~74482. Y.R.\ thanks ELTA for financial support.
\end{acknowledgments}

\appendix

\section*{Appendix: Steady-state solution of the coupled Bloch equations}

In this appendix we derive the steady-state solutions of the Bloch equations in Eq.\,\eqref{Eq: Bloch} for an arbitrary number of species in the vapor cell in response to a monochromatic perturbation. The derivation is inspired by the detailed derivation in the Additional Information in Ref.\,\cite{padniuk2022response}. 

For any real Cartesian vector ${\bf a}$, we define a complex value
\be a_{\perp}\equiv a_x+ia_y \,. \ee
It follows that the transverse complex part of the vectorial product of two vectors is given by
\be ({\bf a}\times{\bf b})_{\perp}=i(-a_{\perp}b_z+a_z b_{\perp}) \ee
and
\be ({\bf a}\times{\bf b})_z=\frac{i}{2}(-a_{\perp}b_{\perp}^*+a_{\perp}^*b_{\perp}) \,. \ee

Working on Eq.\,\eqref{Eq: Bloch}, let us first assume that the applied magnetic field and the pumping direction ${\bf s}$ are in the same direction, $\hat{z}$. Then, in the absence of additional perturbations, all the polarizations are in the same direction, and the steady-state solution ($\dot{P}_j=0$) for the equation reads
\begin{equation}
R_{pj} = \sum_k \left[ \left(\sum_{k'} \kappa_{jk'} n_{k'} + \Gamma_j + R_{pj} \right) \delta_{jk} - \kappa_{jk} n_k \right] P_z^k\,.
\end{equation}
This can be written in a matrix form as
\be \hat{K}P_z=R_p\,, \ee
where $P_z$ and $R_p$ (with no indices) are vectors of length equal to the number of gases in the vapor cell. 
The solution is readily given by
\be P_z=\hat{K}^{-1}R_p\,. \ee

The equations for the transverse components of the polarization are then
\begin{eqnarray} \dot{P}_{\perp}^j &=& \frac{1}{q_j}\left\{i\gamma_j\left[-\bar{b}_{\perp}^jP_z^j-\sum_k \lambda_{jk}M_0^k P_{\perp}^kP_z^j+b_z^jP_{\perp}^j\right.\right. \nonumber \\
&&\left. +\sum_k\lambda_{jk}M_0^k P_z^kP_{\perp}^j\right] +\sum_k \kappa_{jk}n_k(P_{\perp}^k-P_{\perp}^j)-\nonumber \\
&& (\Gamma_j+R_{pj})P_{\perp}^j\Biggr\}\,,
\end{eqnarray}
where $\bar{b}_{\perp}^j\equiv B_{\perp}+L_{\perp}^j+\frac{\xi_j b_{\perp}}{\gamma_j}$ is the perpendicular component of the effective magnetic field acting on species $j$.
This can be written in a matrix form
\be \dot{P}_{\perp}=\hat{A}P_{\perp}+i\frac{\gamma}{q} \bar{b}_{\perp}P_z \,, \ee
where  the last term is a vector where each component is a product of the effective transverse field and the polarization in the longitudinal direction, and the matrix $\hat{A}$ is given by
\begin{eqnarray}
\hat{A}_{jk} &=&
-i\frac{1}{q_j}\Biggl\{\gamma_jP_z^j\lambda_{jk}M_0^k+\kappa_{jk}n_k+ \nonumber \\
&&  \delta_{jk}\left[i\gamma_j\left(b_z^j+\sum_{k}\lambda_{jk}M_0^{k}P_z^{k}\right)- \right. \nonumber \\
&& \left.\left. \sum_{k}\kappa_{jk}n_{k}-\Gamma_j-R_{pj}\right]\right\}\,.
\end{eqnarray}
Let us now take the transverse field perturbation to be monochromatic with frequency $\omega$
\be \bar{b}_{\perp}(t)=\bar{b}_{\perp}(0)e^{i\omega t}\,. \ee
By assuming a similar time dependence of the solution $P_{\perp}$, we obtain
\be i\omega P_{\perp}=\hat{A}P_{\perp}+i\frac{\gamma}{q} \bar{b}_{\perp}P_z\,. \ee
The solution is then readily given by
\be P_{\perp}(\omega)=-i[\hat{A}-i\omega\hat{1}]^{-1}\frac{\gamma}{q} \bar{b}_{\perp}P_z\,. 
\label{eq:P_perp} \ee

Let us now consider a perturbation with frequency $\omega$ in the $y$ direction $\bar{b}_y=\bar{b}_y(0)\cos\omega t$.  This can be written in terms of complex amplitudes as $\bar{b}_{\perp}(t)=\frac{i}{2}\bar{b}_y(0)[e^{i\omega t}+e^{-i\omega t}]$.
The polarization in response to this perturbation is then
\be P_{\perp}(t)=P_+ e^{i\omega t}+ P_- e^{-i\omega t}\,, \ee
where the complex positive and negative frequency components are given in terms of the corresponding field components by Eq.\,(\ref{eq:P_perp}).
The actual polarization vectors are
\begin{eqnarray}
P_x(t) &=& {\rm Re}\{ P\}=\frac12(P_+P_-^*)e^{i\omega t}+{\rm c.c.}\,, \\ 
P_y(t) &=& {\rm Im}\{ P\}=-\frac{i}{2}(P_+-P_-^*)e^{i\omega t}+{\rm c.c.}\,.
\end{eqnarray}

\bibliography{synthetic}

\end{document}